\documentclass[fleqn,usenatbib,useAMS]{mnras}

\usepackage{newtxtext,newtxmath}
\usepackage[T1]{fontenc}
\usepackage{ae,aecompl}

\usepackage{graphicx}	% Including figure files
\usepackage{amsmath}	% Advanced maths commands
\usepackage{amssymb}	% Extra maths symbols

\newcommand{\der}{\textrm{d}}

\title[Black widow evolution]{Black widow evolution:\\magnetic braking by an ablated wind}

\author[S. Ginzburg and E. Quataert]{
Sivan Ginzburg\thanks{E-mail: ginzburg@berkeley.edu}\thanks{51 Pegasi b Fellow.}
and Eliot Quataert
\\
Department of Astronomy and Theoretical Astrophysics Center, University of California, Berkeley, CA 94720, USA}

\date{Accepted XXX. Received YYY; in original form ZZZ}

\pubyear{2015}

\begin{document}
\label{firstpage}
\pagerange{\pageref{firstpage}--\pageref{lastpage}}
\maketitle

\begin{abstract}
Black widows are close binary systems in which a millisecond pulsar is orbited by a companion a few per cent the mass of the sun. It has been suggested that the pulsar's rotationally powered $\gamma$-ray luminosity gradually evaporates the companion, eventually leaving behind an isolated millisecond pulsar. The evaporation efficiency is determined by the temperature $T_{\rm ch}\propto F^{2/3}$ to which the outflow is heated by the flux $F$ on a dynamical time-scale. Evaporation is most efficient for companions that fill their Roche lobes. In this case, the outflow is dominated by a cap around the L1 point with an angle $\theta_g\sim (T_{\rm ch}/T_g)^{1/2}$, and the evaporation time is $t_{\rm evap}=0.46(T_{\rm ch}/T_g)^{-2}\textrm{ Gyr}$, where $T_g>T_{\rm ch}$ is the companion's virial temperature. We apply our model to the observed black widow population, which has increased substantially over the last decade, considering each system's orbital period, companion mass, and pulsar spin-down power. While the original (Fruchter et al. 1988) black widow evaporates its companion on a few Gyr time-scale, direct evaporation on its own is too weak to explain the overall population. We propose instead that the evaporative wind couples to the companion's magnetic field, removes angular momentum from the binary, and maintains stable Roche-lobe overflow. While a stronger wind carries more mass, it also reduces the Alfv\'en radius, making this indirect magnetic braking mechanism less dependent on the flux $t_{\rm mag}\propto t_{\rm evap}^{1/3}$. This reduces the scatter in evolution times of observed systems, thus better explaining the combined black widow and isolated millisecond pulsar populations. 
\end{abstract}


%%%%%%%%%%%%%%%%%%%%%%%%%%%%%%%%%%%%%%%%%%%%%%%%%%

%%%%%%%%%%%%%%%%% BODY OF PAPER %%%%%%%%%%%%%%%%%%

\section{Introduction}\label{sec:introduction}

Black widows are systems with a millisecond pulsar and a low-mass companion, a few per cent the mass of the sun, on a short orbit of several hours. The first such system was detected by \citet{Fruchter1988}, and a few dozen similar ones have been found since. Typically, the companion's orbital period $P_{\rm orb}$ and its minimum mass $m\sin i$ are determined from the periodic timing variations in the pulsar's radio pulse. Up to the last decade, only three black widows were known in the galactic field, with the rest found in globular clusters. Radio followups of {\it Fermi} $\gamma$-ray sources, as well as the designated High Time Resolution Universe survey, have since greatly increased the number of known field millisecond pulsars, a significant fraction of which turned out to be black widows \citep{Ray2012,Keith2013,Roberts2013}. Among these discoveries are the extremely low mass companions to PSR J1719-1438 \citep{Bailes2011} and to PSR J2322-2650 \citep{Spiewak2018}, with minimum masses similar to Jupiter---about 20 times lighter than the companion orbiting the original black widow pulsar.    

According to the prevailing theory \citep[see][for a review and references]{Manchester2017}, millisecond pulsars were spun up to their fast rotation rates by accreting material from a main-sequence binary companion. In this picture, isolated millisecond pulsars are the end products of binary evolution, while black widows represent the missing link, with companions reduced to a fraction of their original mass. 

What is the mechanism that sustains mass loss from black widow companions, eventually leading to their full destruction, leaving behind isolated millisecond pulsars? Gravitational radiation can shrink the binary orbit and drive it towards Roche-lobe overflow. However, by the time the companion is reduced to a few percent of the solar mass, gravitational waves are too weak \citep[e.g. fig. 1 in][]{Romani2016}, leading several authors to consider, instead, the evaporation (ablation) of the companion by high-energy photons or particles powered by the pulsar's spin-down energy (\citealt{Kluzniak1988, Phinney88,Ruderman89a}; cf. \citealt{EichlerLevinson1988,LevinsonEichler1991}).   

The growing number of field black widows discovered in the last decade has motivated a renewed theoretical interest in these systems \citep{Benvenuto2012,Benvenuto2014,Benvenuto2015,Chen2013,JiaLi2015,JiaLi2016,LiuLi2017,Ablimit2019}. Much of the theoretical effort has focused on reproducing the observed black widow population by evolving with time binary systems that initially host a main sequence companion. Despite being a key ingredient in such evolutionary tracks, the evaporation of the companion by the pulsar's irradiation is usually parametrized with a simple linear relation between the pulsar's spin-down power and the companion's mass loss rate, with the coefficient unknown.    

Here, we calculate the evaporation efficiency by modelling the hydrodynamical wind launched off the companion's surface by the incident pulsar radiation. We find that the mass ejection rate does not scale linearly with the pulsar's luminosity, but shows a more complex dependence, and is typically lower than previously assumed. 
We apply our model to the observed black widow population and find that, for the majority of systems, evaporation by the pulsar's radiation on its own is too weak to play a major role in the companion's evolution. As an alternative, we suggest that the evaporative wind may couple to the companion's magnetic field and remove angular momentum from the system, thereby maintaining the companion in stable Roche-lobe overflow. Such a mechanism can greatly amplify the mass loss rate and explain the observations.

The remainder of this paper is organized as follows. In Section \ref{sec:mass_loss} we derive general expressions for the mass loss rate in evaporative winds. In Section \ref{sec:application} we apply these expressions to the observed black widow population and calculate evaporation efficiencies and time-scales. In Section \ref{sec:magnetic} we consider magnetic braking by interaction of the wind with the companion's magnetic field and reanalyse the observations. We summarize and discuss our results in Section \ref{sec:summary}

\section{Evaporative wind}\label{sec:mass_loss}

In this section we derive the rate of mass ablation off the companion's surface by the pulsar's irradiation. We omit order-unity coefficients and focus on the scaling relations. In Section \ref{sec:pressure} we discuss the pressure at which the wind is launched $p_0$. In Sections \ref{sec:flux} and \ref{sec:nozzle} we calculate the outflow rate $\dot{m}$ as a function of $p_0$ and the heating rate per particle $\Gamma$. We discuss $\Gamma$ in Section \ref{sec:heating_rate}. 

\subsection{Wind launching pressure}\label{sec:pressure}

The equilibrium of optically thin gas subject to ionizing radiation has been calculated extensively in the past (see references below). Balancing cooling with heating and ionization with recombination yields an equilibrium gas pressure $p(\rho,F)$, where $\rho$ is the density and $F$ is the radiative flux. Heating and ionization processes scale as $\propto \rho F$ whereas cooling and recombination scale as $\propto \rho^2$; as a result, the equilibrium ionization state and gas temperature $T$ are functions of $\rho/F$ for a given irradiation spectrum. Using the ideal gas law $p=\rho k T/\mu$, with $k$ denoting Boltzmann's constant and $\mu$ the molecular weight, wee see that $p/F$ is a function of $\rho/F$. 

A typical equilibrium curve is given in fig. 2 of \citet[][see also \citealt{BaskoSunyaev73,Krolik81}]{McCrayHatchett75}. At high densities, collisionally excited Hydrogen line cooling, which depends exponentially on the temperature, balances photoionization heating, and maintains an almost constant $T\sim 10^4\textrm{ K}$, such that $p\propto\rho$. At lower densities, Hydrogen becomes ionized and equilibrium between heating (Compton and photoionization of heavy elements) and Bremsstrahlung cooling is achieved by raising $T$. For an approximately constant heating rate per particle, Bremsstrahlung dictates $p\propto\rho T\propto \rho^{-1}$ \citep[e.g.][]{RybickiLightman79}, such that $p(\rho)$ reaches a local minimum value $p_0$.
The temperature and pressure keep rising with decreasing $\rho$ until a maximum temperature $kT_{\rm IC}=\varepsilon/4$ is reached, when inverse Compton scattering balances Compton heating \citep[$\varepsilon$ is the typical photon energy; see][]{RybickiLightman79}. For a more detailed explanation of the various cooling and heating mechanisms and the shape of the equilibrium curve, see \citet{BuffMcCray74}, \citet{London1981}, and chapter 10 in \citet{Krolik1999}.

The pressure and density decrease monotonically when going out in the companion's atmosphere. Once the pressure drops below the local minimum $p_0$, the temperature must jump from $T\sim 10^4\textrm{ K}$ to $T=T_{\rm IC}$ in order to reach equilibrium. However, for a hard enough spectrum, the associated sound speed $c_{\rm IC}=(kT_{\rm IC}/\mu)^{1/2}$ is larger than the escape velocity from the companion $v_g$, launching a hydrodynamic wind \citep{BaskoSunyaev73,McCrayHatchett75,Basko1977}.

We follow \citet{Begelman83} and parametrize the wind launching pressure $p_0$ with the dimensionless
\begin{equation}\label{eq:xi_tag}
\Xi'\equiv\frac{F}{p_0c},    
\end{equation}
where $c$ is the speed of light ($F/c$ is the radiation pressure, which is dynamically unimportant because $F$ is well below the companion's Eddington limit). For a variety of radiation spectra, $\Xi'\sim 1$ (table 2 in \citealt{London1981}; see also \citealt{Krolik81} and \citealt{Krolik1999} but notice the difference in definition between $\Xi'$ and their ionization parameter $\Xi$), and for the remainder of the paper we assume $\Xi'=1$. 

\subsection{Sonic point and mass flux}\label{sec:flux}

We mark the escaping mass flux with $f$. Conservation of momentum sets an upper limit of $f\leq p_0/v_g$, since escaping gas must accelerate to at least the escape velocity $v_g$ by the pressure at the base of the wind $p_0$. This is also the limit derived by \citet{Basko1977} by modelling the outflow as an isothermal \citet{Parker1958} wind. \citet{Ruderman89a,Ruderman89b} adopt a similar value for their nominal mass-loss rate. 
However, as we show below (Fig. \ref{fig:regimes}), black widow systems generally do not quite reach this maximum.

We model the outflow as a non-isothermal wind and write the equations of mass and momentum conservation in spherical symmetry:
\begin{equation}\label{eq:mass_conservation}
\rho v r^2=\textrm{const.}
\end{equation}
\begin{equation}\label{eq:momentum_conservation}
\rho v\frac{\der v}{\der r}=-\frac{\der p}{\der r}-\frac{Gm\rho}{r^2},
\end{equation}
where $m$ is the companion's mass, $r$ is the distance from its centre, $v$ is the outflow velocity, and $G$ is the gravitational constant.
We combine equations \eqref{eq:mass_conservation} and \eqref{eq:momentum_conservation} with the ideal gas law:
\begin{equation}\label{eq:parker_wind}
\frac{\der v^2}{\der r}\left(1-\frac{c_s^2}{v^2}\right)=\frac{4c_s^2}{r}-\frac{2Gm}{r^2}-\frac{2k}{\mu}\frac{\der T}{\der r},   
\end{equation}
with $c_s\equiv(kT/\mu)^{1/2}$ denoting the isothermal sound speed. The initial conditions for equation \eqref{eq:parker_wind} are $v\to 0$, $T\to 0$, and $p=p_0$ at the base of the wind, i.e. on the companion's surface $r=R$.
As usual \citep[e.g.][]{Parker65}, we are seeking for a solution that passes through a sonic point $r_s$, at which $v=c_s$, and accelerates to supersonic velocities. From equation \eqref{eq:parker_wind}, the sonic point is given by \citep{LondonFlannery82}
\begin{equation}\label{eq:sonic_point}
r_s=\frac{Gm\mu}{2kT}+\frac{r_s}{2}\frac{\der\ln T}{\der\ln r}.    
\end{equation}

The first term on the right hand side of equation \eqref{eq:sonic_point} gives the isothermal-wind solution. The temperature profile $\der T/\der r$ in the second term is determined by a competition between the heating and dynamical time-scales of the flow. \citet{Begelman83} characterised this competition with the temperature $T_{\rm ch}$, defined by
\begin{equation}\label{eq:Tch_definition}
kT_{\rm ch}=\Gamma\frac{R}{c_{\rm ch}},
\end{equation}
where $c_{\rm ch}\equiv (kT_{\rm ch}/\mu)^{1/2}$, and $\Gamma$ is the heating rate per particle. Intuitively, this is the temperature reached by the flow on a dynamical time. 
We solve equation \eqref{eq:Tch_definition}:
\begin{equation}\label{eq:Tch_expression}
kT_{\rm ch}=\mu^{1/3}(\Gamma R)^{2/3}\propto F^{2/3}.    
\end{equation}
Since the heating rate is proportional to the incident flux (Section \ref{sec:heating_rate}), $T_{\rm ch}\propto F^{2/3}$ is a measure of the pulsar's irradiation intensity. \citet{Begelman83} identified three wind regimes, which are distinguished by the characteristic temperature $T_{\rm ch}$ and its relation to the Compton temperature $T_{\rm IC}$ and to the virial temperature $T_g\equiv Gm\mu/(kR)\ll T_{\rm IC}$ (no wind is launched if $T_{\rm IC}<T_g$). 
The analytical solution of the equations of motion (including heating) is given by \citet{Begelman83}, while here we provide an intuitive sketch. A summary of the results is provided in Table \ref{tab:regimes}.

\subsubsection{Hot wind: $T_g<T_{\rm IC}<T_{\rm ch}$}\label{sec:hot}

In this regime, the heating time $kT/\Gamma$ is shorter than the dynamical time $R/(kT/\mu)^{1/2}$ even for the maximum temperature $T_{\rm IC}$. The temperature of the flow therefore rises from a low value at the companion's surface, reaches $T_{\rm IC}$ and levels off at a distance $\Delta r\equiv r-R\ll R$. As long as the temperature rises, $\der\ln T/\der\ln r\gg 1$, and there is no solution to equation \eqref{eq:sonic_point}. The sonic point is reached when the asymptotic temperature $T_{\rm IC}$ is approached and $\der\ln T/\der\ln r\approx 2$, at a distance $\Delta r_s\ll R$; the gravitational term in equation \eqref{eq:sonic_point} is negligible $Gm\mu/(kT_{\rm IC})\ll R$. 

At the sonic point, the mass flux is given by $f=\rho c_s=p/c_s$. By integrating equation \eqref{eq:momentum_conservation} we find that $p+\rho v^2$ is conserved as long as the geometry is planar ($\Delta r\ll R)$ and $f=\rho v$ is constant; the gravitational term in the integration is negligible by a factor of $(T_g/T_{\rm IC})(\Delta r/R)\ll 1$. The pressure at the sonic point is therefore given by $p+\rho c_s^2=p_0$, i.e. $p=p_0/2$. We conclude that the mass flux is this regime is $f=p_0/(2c_{\rm IC})$.

\subsubsection{Intermediate: $T_g<T_{\rm ch}<T_{\rm IC}$}\label{sec:intermediate}

We make the ansatz that in this regime the sonic point is located at a distance $\Delta r_s\sim R$ from the surface. From the definition in equation \eqref{eq:Tch_definition}, the temperature there must then be $\sim T_{\rm ch}$. There is no solution to equation \eqref{eq:sonic_point} as long as $\Delta r\ll R$ because $\der\ln T/\der\ln r\gg 1$. When $\Delta r$ becomes comparable to $R$, $\der\ln T/\der\ln r\approx 2$ and the sonic point is reached; the gravitational term in equation \eqref{eq:sonic_point} is negligible in this regime $Gm\mu/(kT_{\rm ch})\ll R$. This justifies our assumption that $\Delta r_s\sim R$.

The geometry is roughly planar for $\Delta r\lesssim R$, so $p+\rho v^2$ is conserved, up to an order-unity correction, by integration of equation \eqref{eq:momentum_conservation} similarly to Section \ref{sec:hot}; the gravitational term in the integration is negligible by a factor of $T_g/T_{\rm ch}$. The pressure at the sonic point therefore satisfies $p+\rho c_s^2\sim p_0$, i.e. $p\sim p_0$. The mass flux through the sonic point is $f=p/c_s\sim p_0/c_{\rm ch}$.

\subsubsection{Cold wind: $T_{\rm ch}<T_g<T_{\rm IC}$}\label{sec:cold}

In this regime, our ansatz that the sonic point is at $\Delta r_s\sim R$---and as a consequence at $T\sim T_{\rm ch}$ (see Section \ref{sec:intermediate})---fails. The gravitational term in equation \eqref{eq:sonic_point} cannot be neglected in this case because $Gm\mu/(kT_{\rm ch})\gg R$, indicating that the sonic point is reached only at a radius $r_s\gg R$. Therefore, at $\Delta r\sim R$ the flow is subsonic ($v\ll c_s$), $\der\ln T/\der\ln r\sim 1$ as before, and the right-hand side of equation \eqref{eq:parker_wind} is negative: $2c_s^2(2-\der\ln T/\der\ln r)-2v_g^2<0$. This expression must change sign at the sonic point, and since $v_g$ decreases with $r$, $c_s\lesssim v_g$ at $\Delta r\sim R$ (similar to an isothermal Parker wind). According to equation \eqref{eq:parker_wind}, solutions with $c_s\ll v_g$ at $\Delta r\sim R$ are characterised by an exponentially fast acceleration on a scale much smaller than $R$: $\der\ln v^2/\der\ln r\sim v_g^2/c_s^2\gg 1$. While these are valid solutions for an isothermal Parker wind, an exponentially subsonic velocity at $\Delta r\sim R$ would give the heated flow in our case enough time $R/v$ to reach temperatures well above $T_g$ (i.e. $c_s\gg v_g$) at $\Delta r\sim R$. We conclude that the only self consistent solution is the one with $c_s\sim v_g$ at $\Delta r\sim R$. 

We find the pressure at $\Delta r\sim R$ by integrating equation \eqref{eq:momentum_conservation} from the surface. The hydrodynamical term is negligible due to the subsonic velocities, and since for $c_s\sim v_g$ the scale height is of order $R$, we find that $p\sim p_0$.

The mass flux at $\Delta r\sim R$ is given by $f=\rho v=(p/c_s)(v/c_s)$. The Mach number $v/c_s$ at $\Delta r\sim R$ is determined by the condition that the subsonic flow is heated to $T_g$ on a time-scale $R/v$: $kT_g=\Gamma R/v$. By comparing this condition to equation \eqref{eq:Tch_definition}, we find that $v/c_{\rm ch}=T_{\rm ch}/T_g$ and $v/c_s\sim (v/c_{\rm ch})(c_{\rm ch}/v_g)=(T_{\rm ch}/T_g)^{3/2}$. Finally, the flux is given by $f\sim (p_0/v_g)(v/c_s)=(p_0/c_{\rm ch})(T_{\rm ch}/T_g)^2$. As also pointed out by \citet{Begelman83}, the mass flux in the cold regime is not determined by conditions at the sonic point, but rather by the Mach number constraint at $\Delta r\sim R$.

\paragraph*{}
The major difference between \citet{Begelman83} and our scenario is the heating function. $\Gamma$ drops as $(r/R)^{-2}$ for the accretion-disc winds of \citet{Begelman83}, whereas in our case $\Gamma$ is constant until $r$ is comparable to the companion's distance from the pulsar $a\gg R$. As discussed above, however, the mass flux in all three wind regimes is determined by conditions at $r-R\lesssim R$, where $\Gamma$ is approximately constant in both cases. The \citet{Begelman83} disc-wind results are therefore applicable to irradiated companions as well.
   
\begin{table}
	\centering
	\caption{Wind regimes identified by \citet{Begelman83} for different values of the characteristic temperature $T_{\rm ch}$ which is defined in equation \eqref{eq:Tch_expression}; the radius of the sonic point is given by $r_s=R+\Delta r_s$.}
	\label{tab:regimes}
	\begin{tabular}{lccc}
		\hline
		 & Hot & Intermediate & Cold\\
		\hline
		Temperature $T_{\rm ch}$ & $>T_{\rm IC}$ & $T_g<T_{\rm ch}<T_{\rm IC}$ & $<T_g$\\
		Sonic point $\Delta r_s$ & $\ll R$ & $\sim R$ & $\gg R$\\
		Mass flux $f$ & $p_0/c_{\rm IC}$ & $p_0/c_{\rm ch}$ & $(T_{\rm ch}/T_g)^2\, p_0/c_{\rm ch}$\\
		\hline
	\end{tabular}
\end{table}

\begin{figure}
\includegraphics[width=\columnwidth]{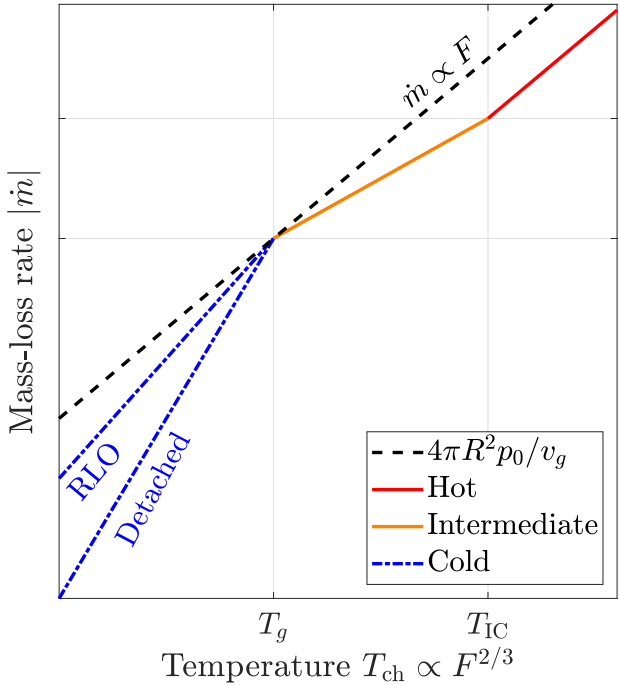}
\caption{Schematic plot of the companion's mass-loss rate $\dot{m}$ as a function of the characteristic temperature $T_{\rm ch}$, which is defined in equation \eqref{eq:Tch_expression}. $T_{\rm ch}$ is a measure of the flux received from the pulsar $F$, and its value compared to the virial ($T_g$) and Compton ($T_{\rm IC}$) temperatures determines the outflow regime (Table \ref{tab:regimes}). The solid red and orange lines depict the hot and intermediate regimes, respectively. Roche-lobe overflowing (RLO) companions differ from detached systems in the cold regime (dot--dashed blue lines; see Section \ref{sec:nozzle}). The dashed black line marks the maximum rate assumed by \citet{Basko1977} and \citet{Ruderman89a,Ruderman89b}. This limit, which is linear in the incident luminosity ($p_0\propto F$; see Section \ref{sec:pressure}), is reached only for a single value $T_{\rm ch}=T_g$.}
	\label{fig:regimes}
\end{figure}

\subsection{Escape through the L1 Lagrange point}\label{sec:nozzle}

In the cold-wind regime (see Table \ref{tab:regimes}), the mass flux is affected by gravity, embodied by the virial temperature $T_g$. We consider two cases with different gravity fields: detached systems, in which the companion is far from filling its Roche lobe, and systems in Roche-lobe overflow.

In detached systems, the gravitational field is spherically symmetric and equal to $g_0=Gm/r^2$. The cold wind mass-loss rate in this case is simply
\begin{equation}\label{eq:detached}
|\dot{m}|\,\text{(detached)}\sim 4\upi R^2\frac{p_0}{c_{\rm ch}}\left(\frac{T_{\rm ch}}{T_g}\right)^2=4\upi R^2\frac{p_0}{v_g}\left(\frac{T_{\rm ch}}{T_g}\right)^{3/2},
\end{equation}
where we have substituted the mass flux from Table \ref{tab:regimes}.

The gravitational field of companions that fill their Roche lobe is weakened by the pulsar's tidal force, especially near the L1 and L2 Lagrange points. The effective gravity component normal to the companion's surface varies approximately as $g_{\rm eff}\approx g_0(1-\cos^2\theta)$, where $\theta$ is measured relative to the line connecting the pulsar's and companion's centres. Close to the L1 point (the L2 point is not irradiated), the effective virial temperature is therefore $T_g^{\rm eff}\approx T_g\theta^2$. Inside a critical angle of
\begin{equation}\label{eq:theta}
\theta_g=\left(\frac{T_{\rm ch}}{T_g}\right)^{1/2}
\end{equation}
the effective gravity is too weak to restrain the flow $T_g^{\rm eff}<T_{\rm ch}$, and the mass flux is given by the intermediate-wind solution $f=p_0/c_{\rm ch}$ (Table \ref{tab:regimes}). The surface area of the $\theta\lesssim\theta_g$ region is $\sim\theta_g^2$ of the total sphere, so the overall mass-loss rate is given by   
\begin{equation}\label{eq:overflowing}
|\dot{m}|\,\text{(overflowing)}\sim 4\upi R^2\theta_g^2\frac{p_0}{c_{\rm ch}}=4\upi R^2\frac{p_0}{v_g}\left(\frac{T_{\rm ch}}{T_g}\right)^{1/2}.
\end{equation}
The cap around L1 dominates the mass flow: while its area is only $T_{\rm ch}/T_g$ of the total sphere, the flux there is higher by a factor of $(T_g/T_{\rm ch})^2$.

Fig. \ref{fig:regimes} shows the three outflow regimes discussed in Table \ref{tab:regimes}. The hot and intermediate regimes are not sensitive to the gravity field and the outflow rate in these cases is simply $|\dot{m}|\sim 4\upi R^2 f$. The cold regime differs for overflowing and detached systems, and is given by equations \eqref{eq:detached} and \eqref{eq:overflowing}. Fig. \ref{fig:regimes} demonstrates that the commonly assumed relation $\dot{m}\propto F$ is usually inaccurate. In fact, the limit $|\dot{m}|\lesssim 4\upi R^2p_0/v_g$ \citep{Basko1977,Ruderman89a,Ruderman89b} is reached only for systems in which $T_{\rm ch}\approx T_g$.

\subsection{Heating rate}\label{sec:heating_rate}

In order to calculate $T_{\rm ch}$ and determine the relevant regime in Fig. \ref{fig:regimes}, we need to estimate the heating rate per particle $\Gamma$, which appears in equation \eqref{eq:Tch_expression}.
Following the discussion in Section \ref{sec:flux}, we are interested in $\Gamma$ for gas temperatures in the range $T_g\lesssim T<T_{\rm IC}$ (even in the cold regime, the temperature reaches $T_g$ at $\Delta r\sim R$; see Section \ref{sec:cold}). The virial temperature of black widow companions is typically $T_g\gtrsim 10^6\textrm{ K}$, for which Compton scattering is the dominant heating mechanism \citep[e.g.][]{BuffMcCray74}.

The Compton heating rate is given by
\begin{equation}\label{eq:heating_rate}
\Gamma\sim\sigma_{\rm T}F\cdot\begin{cases}
x & x\ll1 \\
x^{-1}\ln x & x\gg 1\, ,
\end{cases}
\end{equation}
where $\sigma_{\rm T}$ is the Thomson cross section and $x\equiv \varepsilon/(m_ec^2)$, with $\varepsilon$ denoting the photon energy and $m_e$ the electron mass. When $x\ll 1$, each interacting photon deposits a fraction $\sim x$ of its energy. When $x\gg 1$, photons deposit almost all of their energy, but the cross section is reduced according to the Klein-Nishina formula \citep[e.g.][]{RybickiLightman79}.
Equation \eqref{eq:heating_rate} implies that the heating rate of a roughly flat Crab-like $\gamma$-ray spectrum \citep{BuhlerBlandford2014} is dominated by MeV photons ($\varepsilon\sim m_ec^2$). Furthermore, measurements by {\it Compton} \citep{Kuiper2000} and interpolation of {\it NuSTAR} \citep{GotthelfBogdanov2017} with {\it Fermi} \citep{Abdo2013} data suggest that millisecond pulsar emission might peak around MeV. For these reasons, we estimate that $\Gamma\sim \sigma_{\rm T}F$.

While pair production has a larger cross section than Compton scattering at the highest photon energies, the resulting $e^\pm$ pairs have to deposit their energy high enough in the companion's atmosphere to drive an outflow. A similar challenge is faced by the TeV $e^\pm$ wind that potentially carries a large fraction of the pulsar's spin-down power \citep{Ruderman89a}. According to one scenario, a $\sim 10^2$ G magnetic field may convert the energy of TeV particles into MeV photons by synchrotron radiation \citep{Kluzniak1988,Phinney88,Ruderman89a}. Such secondary photons can be accounted for by adjusting $F$ appropriately. In Section \ref{sec:application} we consider only the direct photons when calculating the ablating flux, possibly underestimating $F$ by a factor of a few. 

\section{Application to observed systems}\label{sec:application}

\begin{table*}
	\centering
	\caption{Black widow sample from the ATNF Pulsar Catalogue  \url{http://www.atnf.csiro.au/research/pulsar/psrcat} \citep{Manchester2005}, version 1.61 (September 2019). We include all systems with pulsar periods $P_{\rm PSR}<10\textrm{ ms}$ \citep[millisecond pulsars are a distinct population, see][]{Lorimer2008,Manchester2017}, minimum companion masses $m\sin i < 7\times 10^{-2}M_\odot$ \citep[distinguishing black widows from the more massive redbacks; see][who discuss this bimodality]{Chen2013,Roberts2013}, and measured spin-down rates $\dot{P}_{\rm PSR}$ from which spin-down luminosities $L_{\rm PSR}$ are inferred \citep[assuming a pulsar moment of inertia $I=10^{45}\textrm{ g cm}^2$; see][]{Manchester2005}. We exclude J1737-0811 because it is a wide binary \citep[$P_{\rm orb}=80\textrm{ days}$;][]{Boyles2013}. Each system's characteristic temperature $T_{\rm ch}$, evaporation efficiency $\eta$, and evaporation time-scale $t_{\rm evap}$ are derived using equations \eqref{eq:t_ch_tg_ratio}--\eqref{eq:time} by assuming a median inclination angle $i=60^{\circ}$, and a $\gamma$-ray luminosity $L_\gamma = 0.1 L_{\rm PSR}$ \citep[fig. 10 in][]{Abdo2013}. The two rightmost columns show each system's magnetic braking time-scale $t_{\rm mag}$, given by equation \eqref{eq:t_mag}, for two choices of the companion's magnetic field $B_0$: a constant $B_0=70\textrm{ G}$ and a variable $B_0=240(P_{\rm orb}/1\textrm{h})^{-0.86}\textrm{ G}$. The sample is sorted by increasing $t_{\rm evap}$}
	\label{tab:sample}
	\begin{tabular}{lcccccccccc}
		\hline
		 PSR Name & $P_{\rm PSR}$ & $m\sin i$ & $P_{\rm orb}$ & $L_{\rm PSR}$ & $T_{\rm ch}/T_g$ & $\eta$ & $P_{\rm PSR}/\dot{P}_{\rm PSR}$ & $t_{\rm evap}$ & $t_{\rm mag}^{B=70}$ & $t_{\rm mag}^{{\rm var}\,B}$\\
		 & (ms) & ($10^{-2}M_\odot$) & (h) & ($L_\odot$) &  & ($10^{-4}$) & (Gyr) & (Gyr) & (Gyr) & (Gyr)\\
		 
		\hline
J1701-3006F    & 2.3    & 2.1    & 4.9    & 190    & 0.88   & 4.5    & 0.33   & 0.58   & 1.9    & 2.3  \\  
J0024-7204P    & 3.6    & 1.7    & 3.5    & 140    & 0.73   & 4.3    & 0.17   & 0.85   & 3.3    & 2.7  \\ 
J1701-3006E    & 3.2    & 3.0    & 3.8    & 92     & 0.44   & 3.9    & 0.33   & 2.4    & 3.8    & 3.4  \\  
J1959+2048     & 1.6    & 2.1    & 9.2    & 41     & 0.36   & 2.4    & 3.0    & 3.4    & 1.7    & 4.3  \\  
J2115+5448     & 2.6    & 2.2    & 3.2    & 44     & 0.30   & 3.1    & 1.1    & 5.0    & 6.3    & 4.7  \\  
J1513-2550     & 2.1    & 1.6    & 4.3    & 23     & 0.24   & 2.3    & 3.1    & 8.1    & 5.8    & 5.9  \\  
J0024-7204R    & 3.5    & 2.6    & 1.6    & 36     & 0.21   & 3.4    & 0.74   & 11     & 17     & 5.6  \\  
J1731-1847     & 2.3    & 3.3    & 7.5    & 20     & 0.18   & 2.1    & 2.9    & 15     & 3.2    & 6.2  \\  
J1311-3430     & 2.6    & 0.82   & 1.6    & 13     & 0.17   & 2.1    & 3.9    & 16     & 26     & 8.3  \\  
J0024-7204O    & 2.6    & 2.2    & 3.3    & 17     & 0.16   & 2.2    & 2.8    & 18     & 9.6    & 7.2  \\  
J1446-4701     & 2.2    & 1.9    & 6.7    & 9.5    & 0.13   & 1.5    & 7.1    & 25     & 5.0    & 8.5  \\  
J1518+0204C    & 2.5    & 3.7    & 2.1    & 17     & 0.11   & 2.6    & 3.0    & 35     & 17     & 7.8  \\  
J2241-5236     & 2.2    & 1.2    & 3.5    & 6.4    & 0.11   & 1.5    & 10     & 37     & 13     & 10   \\     
J2214+3000     & 3.1    & 1.3    & 10     & 4.9    & 0.11   & 1.1    & 6.7    & 37     & 3.9    & 11   \\  
J2234+0944     & 3.6    & 1.5    & 10     & 4.4    & 0.097  & 1.1    & 5.7    & 49     & 4.1    & 11   \\ 
J1641+8049     & 2.0    & 4.0    & 2.2    & 11     & 0.083  & 2.3    & 7.2    & 66     & 20     & 9.5  \\  
J0023+0923     & 3.1    & 1.6    & 3.3    & 4.1    & 0.071  & 1.3    & 8.5    & 91     & 17     & 13   \\     
J1745+1017     & 2.7    & 1.4    & 18     & 1.5    & 0.056  & 0.65   & 31     & 140    & 3.3    & 17   \\ 
J1544+4937     & 2.2    & 1.7    & 2.9    & 3.1    & 0.056  & 1.3    & 23     & 150    & 23     & 15   \\    
J0610-2100     & 3.9    & 2.1    & 6.9    & 2.2    & 0.048  & 0.95   & 9.9    & 200    & 9.3    & 16   \\     
J1719-1438     & 5.8    & 0.11   & 2.2    & 0.41   & 0.045  & 0.51   & 23     & 220    & 67     & 31   \\     
J0636+5129     & 2.9    & 0.69   & 1.6    & 1.5    & 0.045  & 1.0    & 26     & 230    & 63     & 21   \\     
J2322-2650     & 3.5    & 0.074  & 7.8    & 0.14   & 0.036  & 0.26   & 190    & 360    & 21     & 42   \\     
J2017-1614     & 2.3    & 2.6    & 2.3    & 2.0    & 0.033  & 1.2    & 30     & 420    & 38     & 19   \\     
J2051-0827     & 4.5    & 2.7    & 2.4    & 1.4    & 0.026  & 1.1    & 11     & 680    & 43     & 23   \\     
J1836-2354A    & 3.4    & 1.7    & 4.9    & 0.62   & 0.021  & 0.66   & 46     & 1000   & 25     & 29   \\     
		\hline
	\end{tabular}
\end{table*}

Following previous studies \citep{Stevens92,Benvenuto2012,Benvenuto2014,Chen2013,JiaLi2015,JiaLi2016,LiuLi2017}, we define the mass-loss efficiency $\eta$ as
\begin{equation}\label{eq:efficiency}
\frac{Gm\dot{m}}{R}\equiv -\eta L_\gamma\left(\frac{R}{a}\right)^2,    
\end{equation}
where $L_\gamma=4\upi a^2 F$ is the pulsar's $\gamma$-ray luminosity and $a$ is its separation from the companion. $\eta$ measures the fraction of the incident radiation energy that is invested in overcoming the companion's gravity. As discussed in Section \ref{sec:heating_rate}, we only consider the $\gamma$-ray luminosity that directly couples with the companion's upper atmosphere to drive a wind, and not the total spin-down luminosity. We adopt the same beaming factor (i.e. no beaming) as \citet{Abdo2013} for consistency with their $L_\gamma$ measurements.

From here onward, we assume that companions are always close to filling their Roche lobes \citep{Benvenuto2012,Benvenuto2015,Draghis2019}, so $R/a\simeq 0.5(m/M)^{1/3}$ \citep{Eggleton83} where $M$ is the pulsar's mass. Apparently, black widow pulsars occupy the high end of the observed neutron star mass distribution, with $M\gtrsim 2 M_\odot$ \citep[][$M_\odot$ denotes the solar mass]{vanKerwijk2011,Romani2015,Linares2019}, presumably due to preceding mass transfer from their companions. Here, we adopt instead the canonical $M=1.4 M_\odot$, for consistency with previous studies and existing data bases; this difference has a minor effect on our results.
The assumption that the companion is at its maximal (Roche lobe) size enables us to calculate an upper limit of the mass-loss rate as a function of the measured orbital period $P_{\rm orb}$ and mass $m$. Specifically, using equation \eqref{eq:efficiency}, we can relate the loss of orbital energy to the pulsar's luminosity:
\begin{equation}\label{eq:orbital_energy}
\frac{GM\dot{m}}{a}=-0.5^3\eta L_\gamma.
\end{equation}

Several earlier studies either assumed that $\eta\sim 0.1$ \citep{Stevens92,Benvenuto2012} or left it as a free parameter \citep{Chen2013,Benvenuto2014}.\footnote{Originally, \citet{VanDenHeuvel88} introduced $\eta\sim 0.1$ by misciting \citet{Ruderman89b}: both papers define an efficiency $f$, but the definitions differ by a factor of order $v_g/c$, with the more physical \citet{Ruderman89b} estimate being the smaller of the two.} We now apply the theory developed in Section \ref{sec:mass_loss} to evaluate $\eta$ more precisely.
First, we determine the relevant wind regime by calculating the characteristic temperature $T_{\rm ch}$. Using equations \eqref{eq:Tch_expression} and \eqref{eq:heating_rate}
\begin{equation}\label{eq:t_ch_with_gamma}
kT_{\rm ch}=\mu^{1/3}(\sigma_{\rm T}FR)^{2/3},
\end{equation}
and its ratio to the virial temperature is given by
\begin{equation}\label{eq:t_ch_tg_ratio}
\frac{T_{\rm ch}}{T_g}=0.13\left(\frac{L_\gamma}{L_\odot}\right)^{2/3}\left(\frac{m}{10^{-2}M_\odot}\right)^{-4/9}\left(\frac{P_{\rm orb}}{1\,\textrm{h}}\right)^{2/9},
\end{equation}
where we replace $R$ with the effective Roche-lobe radius and scale to typical black widow values ($L_\odot$ is the solar luminosity). Equation \eqref{eq:t_ch_tg_ratio} indicates that black widow systems are typically (see Table \ref{tab:sample}) in the cold wind regime and therefore (see Fig. \ref{fig:regimes}) evaporate somewhat less efficiently than assumed by \citet{Ruderman89a,Ruderman89b}. Next, we substitute $p_0$ from equation \eqref{eq:xi_tag} into equation \eqref{eq:overflowing}, which is appropriate for overflowing systems in the cold regime. Finally, we find the efficiency by comparing to equation \eqref{eq:efficiency}:
\begin{equation}\label{eq:eta_value}
\begin{split}
\eta &\sim\frac{1}{\Xi'}\frac{v_g}{c}\left(\frac{T_{\rm ch}}{T_g}\right)^{1/2}=\\
&=2.2\times 10^{-4}\left(\frac{L_\gamma}{L_\odot}\right)^{1/3}\left(\frac{m}{10^{-2}M_\odot}\right)^{1/9}\left(\frac{P_{\rm orb}}{1\,\textrm{h}}\right)^{-2/9}.
\end{split}
\end{equation}
Black widow evaporation efficiencies are low, so the evaporation time-scales are long: 
\begin{equation}\label{eq:time}
\begin{split}
t_{\rm evap}&=\frac{m}{|\dot{m}|}=\frac{1}{0.5^3\eta}\frac{GMm}{L_\gamma a}=\\
&=27\textrm{ Gyr}\left(\frac{L_\gamma}{L_\odot}\right)^{-4/3}\left(\frac{m}{10^{-2}M_\odot}\right)^{8/9}\left(\frac{P_{\rm orb}}{1\,\textrm{h}}\right)^{-4/9}.    
\end{split}
\end{equation}
By comparing equations \eqref{eq:t_ch_tg_ratio} and \eqref{eq:time}, we find the useful relation
\begin{equation}\label{eq:tevap_ratio}
    t_{\rm evap}=0.46\textrm{ Gyr}\left(\frac{T_{\rm ch}}{T_g}\right)^{-2},
\end{equation}
which indicates that black widow systems are in the cold regime when $t_{\rm evap}>0.46\textrm{ Gyr}$.

\begin{figure}
\includegraphics[width=\columnwidth]{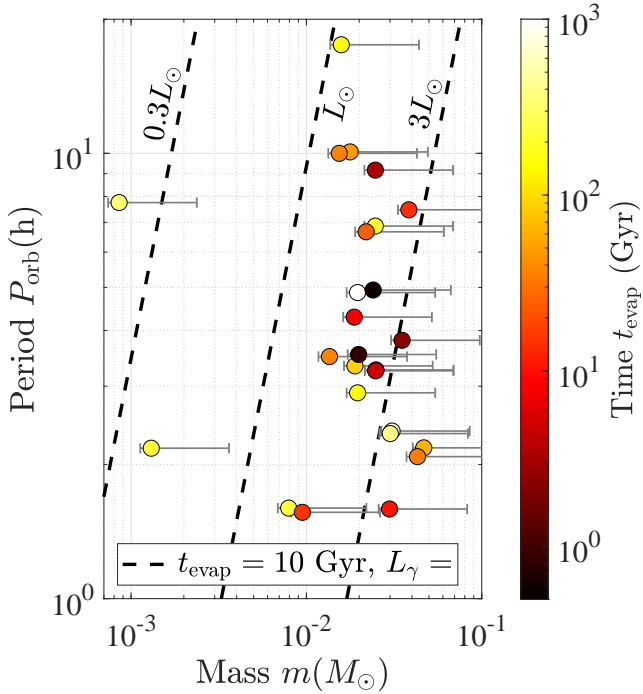}
\caption{The observed black widow sample (Table \ref{tab:sample}). The companion's nominal mass $m$ is calculated assuming the median inclination angle $i=60^\circ$, with the error bars indicating the minimum ($i=90^\circ$) and 95 per cent probability ($i=18.2^\circ$) values. The colour of each system indicates its estimated evaporation time $t_{\rm evap}$, which is also given in Table \ref{tab:sample}. Dark colours correspond to systems in which the pulsar has plausibly evaporated the companion to its low mass over the system's lifetime, whereas light colours represent systems with evaporation times longer than the age of the universe. The dashed black lines are given by equation \eqref{eq:time} with $t_{\rm evap}=10\textrm{ Gyr}$ and different values of the gamma-ray luminosity $L_\gamma$. Evaporation is significant for systems that lie left of the line with the appropriate $L_\gamma$.}
	\label{fig:tevap}
\end{figure}

In Fig. \ref{fig:tevap} we plot $P_{\rm orb}(m,L_\gamma)$ lines (dashed black) for which the evaporation time $t_{\rm evap}=10\textrm{ Gyr}$. Whether or not a pulsar is able to evaporate its companion over a reasonable time-scale depends critically on its $\gamma$-ray luminosity. We populate the figure with the observed black widow sample (Table \ref{tab:sample}), which is limited to pulsars with measured spin-down rates, from which $L_\gamma$ may be estimated. Specifically, \citet{Abdo2013} find that typically $\sim 0.1$ of a millisecond pulsar's total spin-down power $L_{\rm PSR}$ is emitted as $\gamma$ rays in the 0.1--100 GeV {\it Fermi} band (their fig. 10).\footnote{In particular, this relation holds, on average, for the 7 black widow systems in our sample that are also a subset of the \citet{Abdo2013} sample of 40 millisecond pulsars \citep[their table 10; see also 3 additional systems in][which have similar $L_\gamma/L_{\rm PSR}$ fractions]{Barr2013,Bhattacharyya2013,Ray2013}. We have chosen to also include black widows without direct $L_\gamma$ measurements, increasing our sample to a total of 26 systems.} Following the discussion in Section \ref{sec:heating_rate}, we estimate that the MeV photons, which drive the outflow from the companion, carry a comparable share of the luminosity.

Table \ref{tab:sample} and Fig. \ref{fig:tevap} exhibit a three orders of magnitude diversity in estimated evaporation time-scales. Notwithstanding the order-unity uncertainties in our analysis, it is evident that while some black widows may have significantly evaporated their companions by $\gamma$-ray radiation, many others have evaporation time-scales that are longer than the age of the universe. Pulsar spin-down times $P_{\rm PSR}/\dot{P}_{\rm PSR}$, which are typically several Gyrs (Table \ref{tab:sample}), may set an even more stringent constraint.
Interestingly, the original black widow pulsar, PSR J1959+2048 \citep[a.k.a B1957+20; see][]{Fruchter1988}, whose discovery sparked much of the early theoretical work on the subject, has one of the shortest estimated time-scales $t_{\rm evap}=3.4\textrm{ Gyr}$. The two extremely low mass companions to PSR J1719-1438 \citep{Bailes2011} and to PSR J2322-2650 \citep{Spiewak2018}, on the other hand, have two of the longest time-scales, with $t_{\rm evap}=220\textrm{ Gyr}$ and $t_{\rm evap}=360\textrm{ Gyr}$, respectively. Despite the wide range of computed evaporation times $t_{\rm evap}$, all of our systems are in the cold wind regime $T_{\rm ch}<T_g$ (see Table \ref{tab:sample}).  

Even if our analysis underestimates $L_\gamma$ by factors of a few (see Section \ref{sec:heating_rate}), the large scatter in pulsar spin-down luminosities $L_{\rm PSR}$ (Table \ref{tab:sample}) combined with the strong dependence $t_{\rm evap}\propto L_\gamma^{-4/3}$ in equation \eqref{eq:time} indicates that evaporation is unlikely to be the sole driver of black widow evolution. Any correction to $L_\gamma$ or to the evaporation efficiency that brings the longest values of $t_{\rm evap}$ into agreement with the systems' ages would similarly shorten $t_{\rm evap}$ for the other systems and imply a large population of short-lived (sub Gyr) black widows. 
Such short evolution times can be ruled out because isolated field $\gamma$-ray millisecond pulsars are roughly as common as their presumed black widow progenitors, and do not vastly outnumber them \citep{Lorimer2008,Abdo2013}. In Section \ref{sec:magnetic} we propose an alternative mechanism for the evolution of black widows that resolves this puzzle. 

\begin{figure}
\includegraphics[width=\columnwidth]{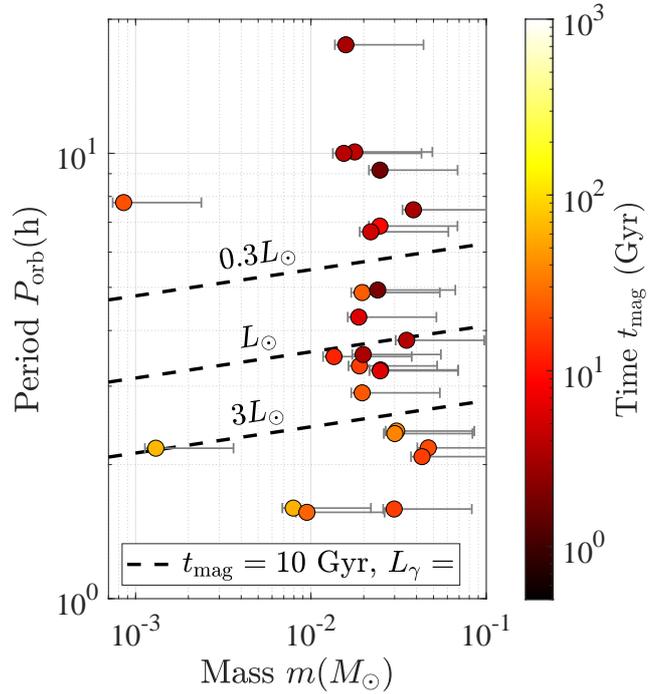}
\caption{Same as Fig. \ref{fig:tevap}, but showing the magnetic braking time-scale $t_{\rm mag}$, given by equation \eqref{eq:t_mag}, assuming a companion magnetic field of $B_0=70$ G ($t_{\rm mag}$ is also given in the second from the right column in Table \ref{tab:sample}). Compared to Fig. \ref{fig:tevap}, evolution times are typically shorter, and more importantly, have a much smaller scatter (i.e. similar colour).}
	\label{fig:tmag}
\end{figure}

\begin{figure}
\includegraphics[width=\columnwidth]{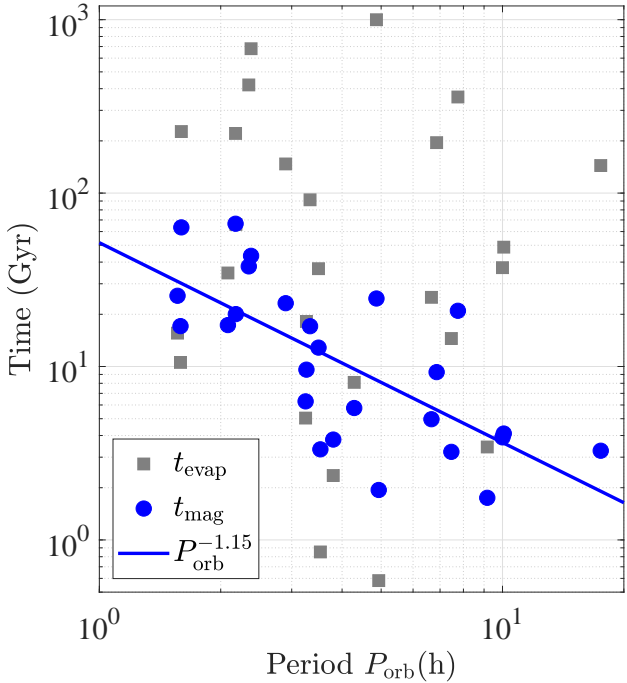}
\caption{Each system's evaporation (grey squares, taken from Fig. \ref{fig:tevap}) and magnetic braking (with a constant $B_0=70$ G; blue circles, taken from Fig. \ref{fig:tmag}) time-scales as a function of orbital period. The magnetic braking times and orbital periods are correlated, with a best fit of $t_{\rm mag}\propto P_{\rm orb}^{-1.15}$ (solid blue line). A variable magnetic field that increases with the rotation rate as $B_0\propto\Omega^{0.86}\propto P_{\rm orb}^{-0.86}$ can remove this correlation (rightmost column in Table \ref{tab:sample}, best fitted by a horizontal $t_{\rm mag}=10\textrm{ Gyr}$).}
	\label{fig:scatter}
\end{figure}

\section{Magnetic braking}\label{sec:magnetic}

In Section \ref{sec:application} we demonstrated that evaporation alone has difficulties in explaining how black widow companions lose a significant fraction of their mass during the pulsar's lifetime. In principle, any process that removes angular momentum from a binary can keep it in stable Roche-lobe overflow \citep{Rappaport1982}, providing an alternative to evaporation.\footnote{Roche-lobe overflow is stable when $5/3+\xi-2\alpha>0$ \citep[e.g.][and references therein]{Rappaport1982,GinzburgSari2017}, where $\xi$ parametrizes the companion's mass-radius relation $R\propto m^\xi$, and $\alpha$ is the fraction of the companion's specific orbital angular momentum that is lost through overflow ($1-\alpha$ presumably returns to the orbit by torques from the disc that forms around the pulsar). $\xi=-1/3$ for both fully degenerate cold companions and for hot companions that lose mass adiabatically, so we estimate that the overflow is stable as long as at least a third of the angular momentum is conserved. Unstable Roche-lobe overflow would lead to fast destruction of companions on dynamical time-scales; this scenario seems to be at odds with the observed black widow population \citep{Draghis2019}.} As discussed in the introduction, gravitational wave emission is too slow, so we must look for a different mechanism. 

A well known mechanism to remove angular momentum from a magnetized spinning object is magnetic braking by an outflow which interacts with the co-rotating magnetic field \citep{WeberDavis67}. In our case, the companion's spin is tidally locked to its orbit so magnetic braking taps into the orbital angular momentum. Previous studies \citep{Chen2013,Benvenuto2014} implemented magnetic braking according to the \citet{Rappaport1983} prescription. This prescription, as explicitly stated by \citet{Rappaport1983}, was not derived from first principles but rather tailored to fit observed rotation velocities of $\sim$$1 M_\odot$ main sequence stars. While this braking law might be relevant at earlier stages of their evolution, it cannot be directly extrapolated to $\sim$$10^{-2} M_\odot$ black widow companions. Moreover, whereas main sequence stars power their own winds, the outflows from black widow companions are driven by incident pulsar irradiation (Section \ref{sec:mass_loss}), and therefore remove mass and angular momentum at a very different rate. 
Here we estimate the magnetic braking rate by coupling the companion's mass loss $\dot{m}$ (Section \ref{sec:mass_loss}) to its magnetic field, for which we assume a surface value of $B_0$. Our calculation closely follows \citet{Thompson2004}, where more details can be found. 

We assume a split monopole geometry \citep{WeberDavis67, Thompson2004} for which the magnetic field decays with radius as
\begin{equation}\label{eq:B}
B(r)=B_0\left(\frac{R}{r}\right)^2.    
\end{equation}
The escaping mass is forced to co-rotate with the magnetic field until it reaches the Alfv\'en radius $R_{\rm A}$, where the kinetic energy density is comparable to the magnetic energy density $\rho v^2/2=B^2/(8\upi)$. Using equation \eqref{eq:B} and $|\dot{m}|=4\upi r^2\rho v$, $R_{\rm A}$ is given by
\begin{equation}\label{eq:RAvel}
B_0^2R_{\rm A}^2\left(\frac{R}{R_{\rm A}}\right)^4=|\dot{m}|v\left(R_{\rm A}\right).
\end{equation}
Spherical symmetry is applicable when deriving equation \eqref{eq:RAvel} even in the asymmetric Roche-lobe overflowing case (Section \ref{sec:nozzle}). Although the outflow from the companion's surface is dominated by a cap with a polar angle $\theta_g<1$ in this case, we find below that $R_{\rm A}>a\gg R$, allowing the outflow to symmetrize by the time it reaches $R_{\rm A}$. 
On the other hand, in more realistic geometries, equatorial magnetic field lines form closed loops \citep[e.g. fig. 1 in][]{MestelSpruit1987} and are thus unable to shuttle angular momentum to large distances. In this scenario, only the spherically symmetric component of the wind, which is launched from higher latitudes and constitutes only a fraction $T_{\rm ch}/T_g$ of the total outflow, contributes to magnetic braking. Since we find below that the magnetic braking time-scale $t_{\rm mag}\propto \dot{m}^{-1/3}$ depends only weakly on $\dot{m}$, our results are insensitive to this effect and are given below assuming the simple split monopole geometry.\footnote{Disregarding the equatorial L1 cap would slightly alter the scaling relations of equation \eqref{eq:t_mag} to $t_{\rm mag}\propto L_\gamma^{-2/3}P_{\rm orb}^{-4/3}m^{2/9}B_0^{-4/3}$ and would require a somewhat stronger magnetic field $B_0=140\textrm{ G}$ to reach the same median $t_{\rm mag}=10\textrm{ Gyr}$ for our sample.}
The Alfv\'en radius marks the transition from a magnetically dominated flow, which rotates with the companion and its magnetosphere, to a free radial outflow. Therefore, at the transition $v\left(R_{\rm A}\right)\sim \Omega R_{\rm A}$, where $\Omega=2\upi/P_{\rm orb}$ is the spin (and through tidal locking also the orbital) angular velocity. As pointed out by \citet{Thompson2004}, this relation applies only if $\Omega R_{\rm A}$ is larger than the outflow velocity from a non-spinning companion, which is equal to the escape velocity $v_g$ in the cold wind regime \citep{Begelman83}. Using the same hierarchy as above $\Omega R_{\rm A}\gg \Omega R\sim v_g$, where the last similarity holds for companions that fill their Roche lobe. 
Substituting for $v\left(R_{\rm A}\right)$ in equation \eqref{eq:RAvel}:
\begin{equation}\label{eq:RA}
\left(\frac{R_{\rm A}}{a}\right)^3=\left(\frac{R}{a}\right)^3\left(\frac{R_{\rm A}}{R}\right)^3=\frac{0.5^3}{2\upi}\frac{B_0^2RP_{\rm orb}}{M}\frac{m}{|\dot{m}|}.
\end{equation}

Since the wind co-rotates with the companion up to $R_{\rm A}$, it carries away angular momentum at a rate $\dot{J}=\dot{m}\Omega R_{\rm A}^2$. The total angular momentum of the system is dominated by the orbit $J=m\Omega a^2$, so the magnetic braking time-scale is 
\begin{equation}\label{eq:t_mag}
\begin{split}
t_{\rm mag}&=\frac{J}{|\dot{J}|}=\frac{m}{|\dot{m}|}\left(\frac{a}{R_{\rm A}}\right)^2=\frac{(2\upi)^{2/3}}{0.5^2}t_{\rm evap}^{1/3}\left(\frac{M}{B_0^2RP_{\rm orb}}\right)^{2/3}=\\
&=31L_\gamma^{-4/9}P_{\rm orb}^{-34/27}\left(\frac{m}{10^{-2}M_\odot}\right)^{2/27}\left(\frac{B_0}{10^2\,\textrm{G}}\right)^{-4/3}\textrm{ Gyr},
\end{split}
\end{equation}
where we have substituted $t_{\rm evap}$ from equation \eqref{eq:time}. In the bottom line of equation \eqref{eq:t_mag}, $L_\gamma$ is measured in units of $L_\odot$ and $P_{\rm orb}$ in hours. Equation \eqref{eq:t_mag} indicates that in order for magnetic braking to play an important role (i.e. $t_{\rm mag}<t_{\rm evap}$), the magnetic field must be strong enough to satisfy $R_{\rm A}>a$, justifying our derivation of equation \eqref{eq:RA}. 

If $t_{\rm mag}<t_{\rm evap}$, the system evolves, and the companion loses mass, primarily as a result of magnetic braking which keeps the companion in stable Roche-lobe overflow, rather than directly by photoevaporation. Mass is transferred to the pulsar through the L1 point at a very narrow angle, which is related to the ratio of the companion's dynamical time to the system's age \citep{LinialSari2017}. The rate of magnetic braking is a function of the evaporative outflow, which is launched at a much larger angle $\theta_g$ and carries mass and angular momentum to a distance of $R_{\rm A}$. Through this process, the pulsar's radiation and the wind that it launches affect the system's evolution indirectly.  
The pulsar may reject---due to its rapid spin or strong wind---the mass transferred to it through Roche-lobe overflow, and eject it to large distances in a `propeller' mechanism \citep{IllarionovSunyaev75}. Such mass ejection removes angular momentum from the pulsar's spin, but not from the binary orbit, from which the pulsar is tidally decoupled. Consequently, magnetic braking is powered only by the mass lost directly in the evaporative wind, even when this is outweighed by the Roche-lobe overflow (in stable Roche-lobe overflow, the angular momentum loss rate is, by definition, set by external mechanisms, such as gravitational wave emission or magnetic braking, and not by the Roche-lobe overflow itself). Accordingly, in our notation $\dot{m}$ refers only to the wind, rather than the total, mass loss rate. 

\subsection{Reanalysis of the observations}

The magnetic braking time-scale depends weakly on $\dot{m}$ as a result of competing effects: a stronger wind carries away more mass, but also constricts the magnetosphere according to equation \eqref{eq:RA}, $R_{\rm A}\propto \dot{m}^{-1/3}$, reducing the specific angular momentum lost. From equation \eqref{eq:t_mag}, $t_{\rm mag}\propto \dot{m}^{-1}R_{\rm A}^{-2}\propto\dot{m}^{-1/3}$, which is a much shallower function of $\dot{m}$ than direct evaporation $t_{\rm evap}\propto\dot{m}^{-1}$. This weak dependence on $\dot{m}$, and by extension on the wind-driving luminosity $L_\gamma$, may help to reduce the scatter in characteristic time-scales inferred for observed systems (Fig. \ref{fig:tevap}). 

In Fig. \ref{fig:tmag} we re-plot the observed black widow sample, this time coloured according to the magnetic braking time-scale $t_{\rm mag}$, assuming the same magnetic field of $B_0=70$ G for all companions. This value was chosen so that the median $t_{\rm mag}=10\textrm{ Gyr}$. Estimating the magnetic field strength of black widow companions from first principles is beyond the scope of this work. Nonetheless, we note that Hot Jupiters have similar radii and exhibit fields similar in strength to our nominal $B_0$ \citep{YadavThorngren2017,Cauley2019}. Typical black widow companions are more massive, rotate faster, and are subject to stronger irradiation---all potentially increase $B_0$ \citep{Christensen2010}. In addition, as briefly mentioned in Section \ref{sec:heating_rate}, $10^2$ G companion fields were historically invoked as a mechanism to produce MeV photons through synchrotron radiation \citep{Kluzniak1988,Ruderman89a}.

As anticipated, Fig. \ref{fig:tmag} displays a significantly smaller scatter in time-scales compared to Fig. \ref{fig:tevap}. This result indicates that magnetic braking, rather than direct evaporation, is a more likely candidate for the dominant evolutionary process that sculpts the observed black widow population.\footnote{Systems evolve on the shorter of the two time-scales. For several systems, $t_{\rm evap}\lesssim t_{\rm mag}$ by factors of a few (or even less), but in most cases, $t_{\rm mag}<t_{\rm evap}$ (Table \ref{tab:sample}).} Moreover, whereas the scatter of evaporation times in Fig. \ref{fig:tevap} does not seem to correlate with either mass or orbital period, Fig. \ref{fig:tmag} shows some correlation between $t_{\rm mag}$ and the orbital period: systems on shorter orbits seem to evolve more slowly. The correlation is easier to notice in Fig. \ref{fig:scatter}, where we plot both time-scales as a function of $P_{\rm orb}$. A possible explanation of this trend is that the magnetic field $B_0$ is not constant, as we have assumed, but in fact increases with the companion's rotation rate $\Omega$ (which is tidally locked to its orbital motion). From equation \eqref{eq:t_mag}, a relation $B_0\propto\Omega^{0.86}$ can remove the trend in Fig. \ref{fig:scatter} and infer a population with a reasonable scatter, given the order-unity uncertainties of our analysis, around a median lifetime of 10 Gyr (rightmost column in Table \ref{tab:sample}). Such a positive correlation between $B_0$ and $\Omega$ is in accord with many (but not all) dynamo scaling laws that have been suggested in the literature \citep[see][for a review; specifically their table 1]{Christensen2010}.

\section{Summary and discussion}\label{sec:summary}

The last decade has seen a substantial increase in the population of observed black widow pulsars, with some systems very different from the original one detected by \citet{Fruchter1988}. The growing number of black widows may allow us to statistically check which mechanisms control their formation and evolution. One mechanism, which was suggested early on, is the evaporation (ablation) of the low-mass companion by the high energy radiation of its host pulsar \citep{Kluzniak1988,Phinney88}. In recent years, evaporation has been implemented in binary evolution codes with a simple prescription, which relates the evaporation rate to the pulsar's spin-down luminosity, introducing an unknown efficiency parameter \citep{Benvenuto2012,Chen2013}.  

Here, we calculated the evaporation efficiency by studying the hydrodynamical \citet{Parker1958} wind launched off the companion's atmosphere. The wind's structure can be solved by considering the interplay between thermal and dynamical processes, encapsulated by the characteristic temperature $T_{\rm ch}\propto F^{2/3}$ which can be reached in a sound crossing time given the incident flux $F$ \citep{Begelman83}. Black widows are in the `cold wind' regime, characterised by $T_{\rm ch}<T_g$ (the companion's virial temperature). In this regime, gravity retards the flow and the mass loss rate does not reach the maximum value assumed by \citet{Basko1977} and \citet{Ruderman89a,Ruderman89b}. The pulsar's tidal force weakens the effective surface gravity on companions that fill their Roche lobe. The outflow in this case is stronger when compared to detached companions (but still below the maximum; see Fig. \ref{fig:regimes}), and it is dominated by a cap around the L1 Lagrange point with a polar angle $\theta_g\sim(T_{\rm ch}/T_g)^{1/2}$. The wind launched from this cap is in the `intermediate' regime (Table \ref{tab:regimes}). The time-scale to fully evaporate the companion is $t_{\rm evap}=0.46(T_{\rm ch}/T_g)^{-2}\textrm{ Gyr}$ (equation \ref{eq:tevap_ratio}).  

We applied our model to a sample of 26 black widows, considering each system's orbital period, companion mass, and pulsar spin-down luminosity. We put a lower limit on $t_{\rm evap}$ by assuming that companions fill their Roche lobe and maximize their radius, consistent with recent observational constraints \citep{Draghis2019}. Even with this assumption, most companions evaporate on a time-scale that is longer than either the age of the universe or the pulsar's spin-down time (Table \ref{tab:sample}).
This conundrum cannot be resolved by a simple multiplicative correction: the large dispersion in measured spin-down luminosities, combined with the strong dependence on the incident flux $t_{\rm evap}\propto F^{-4/3}$, leads to a three orders of magnitude scatter in evaporation time-scales, with no apparent trend (Fig. \ref{fig:tevap}). Any arbitrary enhancement of the evaporation efficiency that expedites the evolution of systems with the longest $t_{\rm evap}$, or even assuming that the spin-down luminosity directly couples to the companion with 100 per cent efficiency, would also shorten the shortest $t_{\rm evap}$, implying a large population of short lived black widows that quickly destroy their companions. Such a population is inconsistent with the relative numbers of observed black widows and isolated millisecond pulsars. We conclude that evaporation on its own is not responsible for the evolution of most black widows and their eventual transformation into isolated millisecond pulsars. 

Although the pulsar's irradiation is too weak to directly evaporate the companion on a relevant time-scale, the evaporative wind may couple to the companion's magnetic field and remove angular momentum from the binary system. This magnetic braking can keep the companion in stable Roche-lobe overflow and reduce its mass at a much faster rate. While a stronger wind removes more mass, it also decreases the Alfv\'en radius and by extension the specific angular momentum carried away. These competing effects lead to a weak dependence of the magnetic braking time-scale on the wind strength and on the flux $t_{\rm mag}\propto t_{\rm evap}^{1/3}\propto F^{-4/9}$, significantly reducing the scatter in evolution times (Fig. \ref{fig:tmag}). Quantitatively, a companion field of about $B_0\sim 70\textrm{ G}$ can explain how the observed population evolves on a time-scale of $\sim 10\textrm{ Gyr}$; the scatter can be further minimized if the magnetic field increases with the companion's spin rate as $B_0\propto\Omega^{0.86}$. The shorter time-scales and smaller scatter are more compatible with the combined population of observed black widows and isolated millisecond pulsars.

Previous studies tried to directly constrain the evaporative wind $\dot{m}$ by fitting an `intra-binary shock' model to optical observations of black widow companions \citep{RomaniSanchez2016}. The idea is that the collision between winds emanating from the pulsar and its companion forms a shock which reprocesses the pulsar's spin-down luminosity and heats the companion non-uniformly. In principle, the shape of the shock, and hence the relative strength of the two winds, can be deduced from the modulation in the companion's light curve. In practice, however, \citet{RomaniSanchez2016} find a weak dependence on $\dot{m}$ which is degenerate with the other parameters of their model. Recently, \citet{Draghis2019} fit a direct heating model (disregarding the intra-binary shock) to the optical light curves of nine black widow systems.
They find that eight of the systems fill or almost fill their Roche lobes, with (radius) filling factors of 0.7--1 (their table 3, the ninth system has a filling factor of 0.5). Given the uncertainties of the fitting model, it is plausible that all observed systems completely fill their Roche lobes, as also implied by our combined magnetic braking and Roche-lobe overflow scenario. 

\section*{Acknowledgements}

We thank Jeremy Hare, Adam Jermyn, Sterl Phinney, Roger Romani, Re'em Sari, and Ken Shen for discussions. SG is supported by the Heising-Simons Foundation through a 51 Pegasi b Fellowship.  

%%%%%%%%%%%%%%%%%%%%%%%%%%%%%%%%%%%%%%%%%%%%%%%%%%



\bibliographystyle{mnras}
%\bibliography{black} % if your bibtex file is called example.bib
\input{black.bbl}

%%%%%%%%%%%%%%%%%%%%%%%%%%%%%%%%%%%%%%%%%%%%%%%%%%

%%%%%%%%%%%%%%%%% APPENDICES %%%%%%%%%%%%%%%%%%%%%

%%%%%%%%%%%%%%%%%%%%%%%%%%%%%%%%%%%%%%%%%%%%%%%%%%

% Don't change these lines
\bsp	% typesetting comment
\label{lastpage}
\end{document}